\documentclass[12pt,letterpaper]{article}
\usepackage{epsfig,graphicx,setspace}
\usepackage{amsmath,amsthm,amssymb,enumerate}
\usepackage{bm,verbatim}
\usepackage{enumitem}
\usepackage{amsfonts}
\usepackage{color}
\usepackage{hyperref}
\usepackage[flushleft]{threeparttable}
\usepackage{booktabs,caption}
\usepackage{rotating}
\usepackage{float}
\usepackage{subcaption}
\usepackage{multirow}
\usepackage{cite}
\usepackage{makecell}
\usepackage[T1]{fontenc}
\usepackage{indentfirst}
\usepackage{lscape}

\newcommand{\RNum}[1]{\uppercase\expandafter{\romannumeral #1\relax}}
\long\def\symbolfootnote[#1]#2{\begingroup\def\thefootnote{\fnsymbol{footnote}}
	\footnote[#1]{#2}\endgroup}

\usepackage{hyperref}
\hypersetup{hidelinks,
	colorlinks=true,
	allcolors=black,
	pdfstartview=Fit,
	breaklinks=true}

\begin{document}
\begin{center}
	\section*{Semi-Supervised Hybrid Spine Network for Segmentation of Spine MR Images}
	Meiyan Huang, Shuoling Zhou, Xiumei Chen, Haoran Lai, Qianjin Feng
\end{center}



\begin{abstract}
Automatic segmentation of vertebral bodies (VBs) and intervertebral discs (IVDs) in 3D magnetic resonance (MR) images is vital in diagnosing and treating spinal diseases. However, segmenting the VBs and IVDs simultaneously is not trivial. Moreover, problems exist, including blurry segmentation caused by anisotropy resolution, high computational cost, inter-class similarity and intra-class variability, and data imbalances. We proposed a two-stage algorithm, named semi-supervised hybrid spine network (SSHSNet), to address these problems by achieving accurate simultaneous VB and IVD segmentation. In the first stage, we constructed a 2D semi-supervised DeepLabv3+ by using cross pseudo supervision to obtain intra-slice features and coarse segmentation. In the second stage, a 3D full-resolution patch-based DeepLabv3+ was built. This model can be used to extract inter-slice information and combine the coarse segmentation and intra-slice features provided from the first stage. Moreover, a cross tri-attention module was applied to compensate for the loss of inter-slice and intra-slice information separately generated from 2D and 3D networks, thereby improving feature representation ability and achieving satisfactory segmentation results. The proposed SSHSNet was validated on a publicly available spine MR image dataset, and remarkable segmentation performance was achieved. Moreover, results show that the proposed method has great potential in dealing with the data imbalance problem. Based on previous reports, few studies have incorporated a semi-supervised learning strategy with a cross attention mechanism for spine segmentation. Therefore, the proposed method may provide a useful tool for spine segmentation and aid clinically in spinal disease diagnoses and treatments. Codes are publicly available at: https://github.com/Meiyan88/SSHSNet.
\end{abstract}

Keywords: Spine segmentation, MR images, semi-supervised learning, cross tri-attention.

\section{Introduction}
\label{sec:introduction}
\begin{figure}
  \centering
  \includegraphics[width=8.8cm]{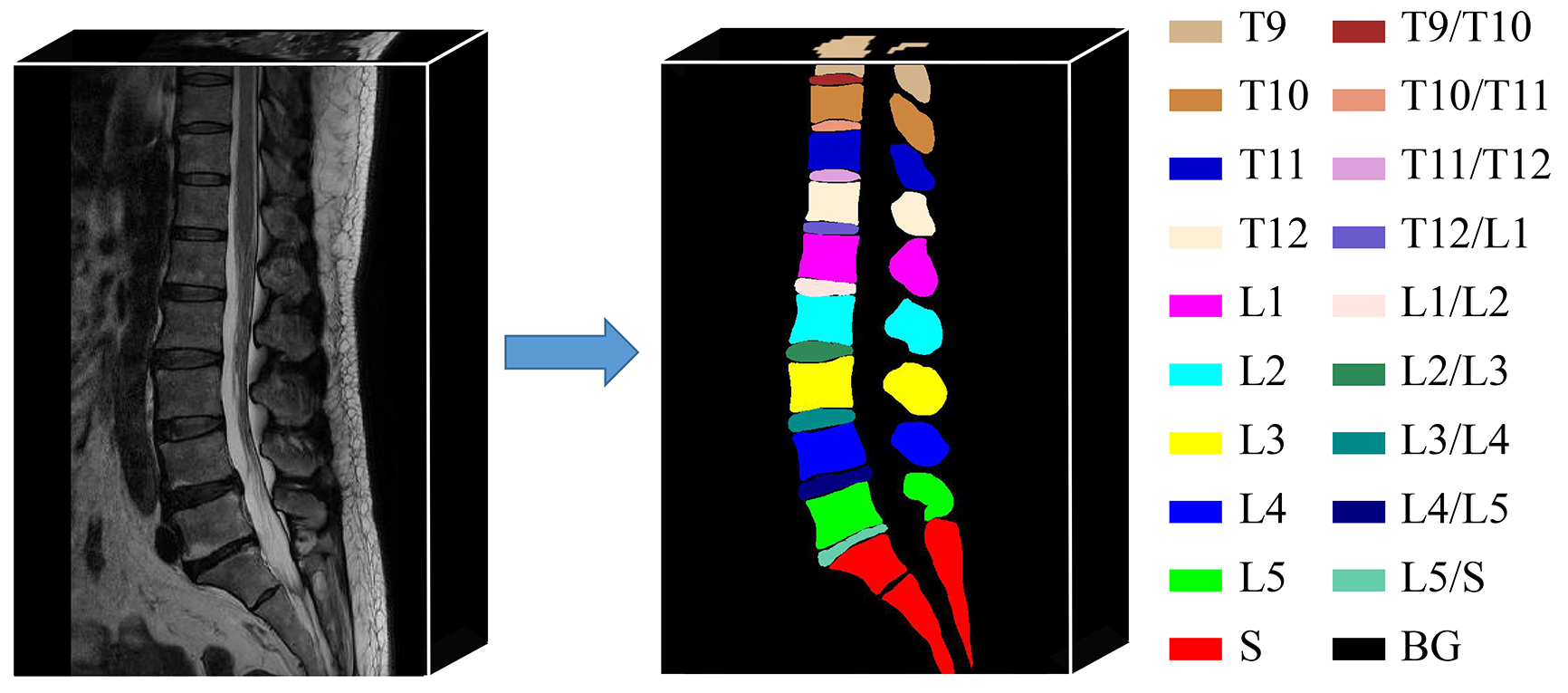}
  \caption{ An example of spine segmentation in MR image. This image has 20 labels. “BG” represents background. “T”, “L”, and “S” represent thoracic, lumbar, and sacral VBs, respectively. Moreover, -/- represents IVD.}
  \label{fig1}
\end{figure}

Abstract{T}{he} spine, where vertebral bodies (VBs) and intervertebral discs (IVDs) share most of the body weights, is an indispensable part of the human body structure \cite{r1}. However, the VBs and IVDs undergo morphological changes because of the increase in age and decrease in physical activity, leading to impaired spinal function (i.e., spinal degenerative disease). Nowadays, spine segmentation (Fig. \ref{fig1}), which is referred to as multi-class segmentation of the VBs and IVDs for spine MR images \cite{r2}, can be employed in spinal disease diagnosis \cite{r3}, surgical treatment planning \cite{r4}, and spine pathology identification \cite{r3}. However, manual spine segmentation is time consuming and depends on the individual operator. Therefore, designing an automatic spine segmentation approach is necessary to provide acceptable performance for clinical application.

\begin{figure}
  \centering
  \includegraphics[width=8.8cm]{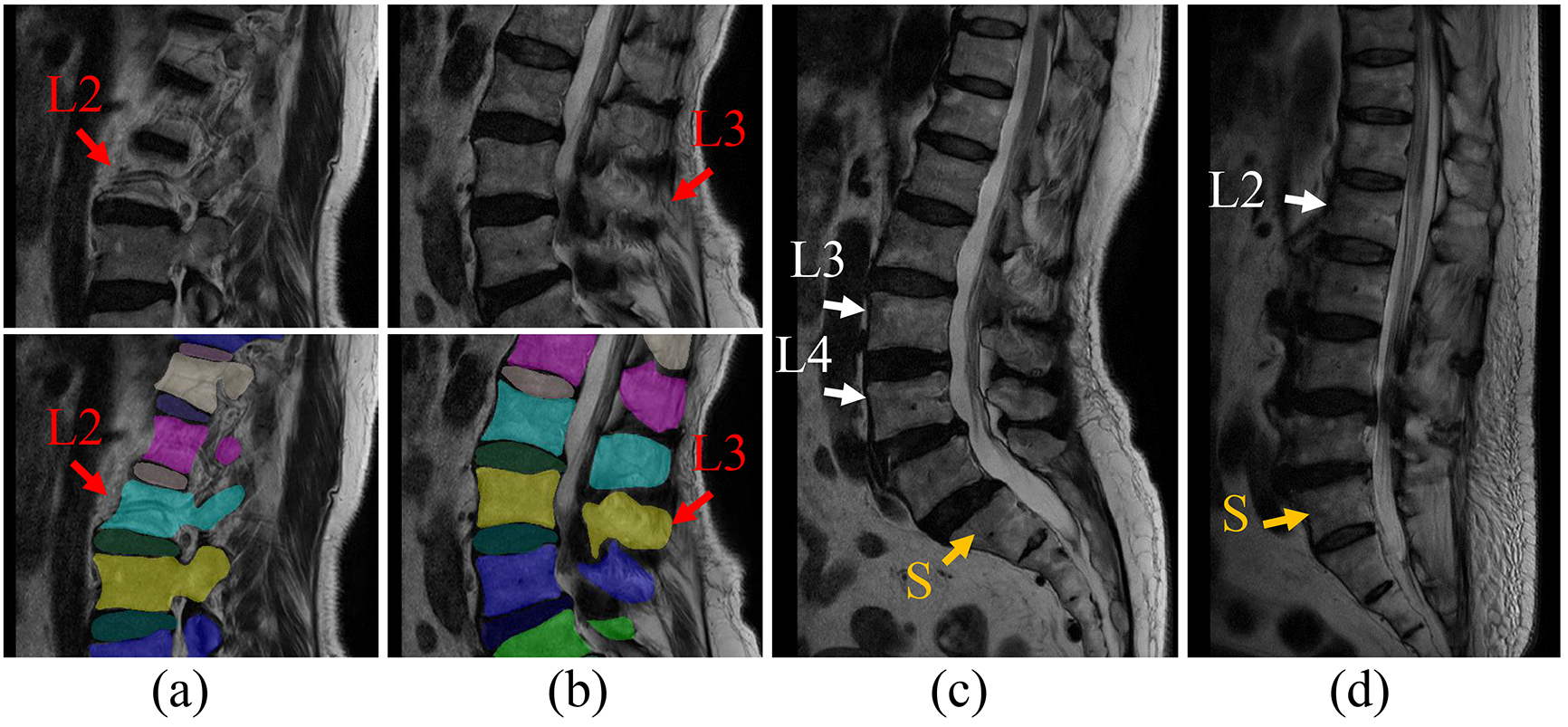}
  \caption{The challenges for spine segmentation in MR images. (a) Intensity inhomogeneity between edge and center in L2 (indicated by red arrows). (b) Partial volume effect results in the blurred edge of the vertebral arch of L3 (indicated by red arrows). The spine MR images (c) without and (d) with lumbarization. Inter-class similarities appear within a sample (L3 and L4 in (c), indicated by white arrows) and among different samples (L3 in (c) and L2 in (d), indicated by white arrows), whereas a different appearance of the sacral cone is observed in different samples (indicated by yellow arrows in (c) and (d)).}
  \label{fig2}
\end{figure}

Spine segmentation in volumetric MR images is vital. However, several challenges exist during the segmentation. First, anisotropic spatial resolution exists in spine MR images, resulting in intensity inhomogeneity (Fig. \ref{fig2} (a)) and partial volume effect (Fig. \ref{fig2} (b)). Thus, segmentation difficulty increases. Second, inter-class similarity and intra-class variety appear in spine MR images. As illustrated in Figs. \ref{fig2} (c) and \ref{fig2} (d), inter-class similarities exist within a sample and among different samples, whereas intra-class varieties appear in samples with/without lumbarization. Third, high computational memory caused by high dimensional images exists. Fourth, two types of data imbalances exist in spine segmentation: imbalance of data labels and imbalance of sample types. Data label imbalance is embodied in the sample number with T9 and T9/T10, which is much lower than the sample number without T9 and T9/T10. In sample type imbalance, the sample number with lumbar sacralization or lumbarization is a small proportion of the entire data. This scenario may easily lead to misprediction between sacral and L5.

A two-stage method, namely, semi-supervised hybrid spine network (SSHSNet), was proposed in this study to overcome the aforementioned challenges by achieving accurate spine segmentation in volumetric MR images. In the first stage, we used compressed but not cropped 2D slices as inputs of a 2D semi-supervised network to obtain the intra-slice information containing relationships between adjacent connections of VBs and IVDs. In particular, cross pseudo supervision (CPS) \cite{r5}, a kind of semi-supervised learning method, was introduced to achieve stable and coarse segmentation. The outputs in this stage can be used as prior knowledge to guide fine segmentation in the next stage. In the second stage, we used the full-resolution but cropped 3D mini-patches and coarse segmentations generated from the first stage as inputs of a 3D fine segmentation network to obtain the inter-slice information containing locally spatial information. Moreover, a cross tri-attention module (CTAM) was proposed to fuse the information extracted from these two networks effectively. Therefore, comprehensive intra- and inter-slice information can be well exploited by using the proposed method for the simultaneous segmentation of the VBs and IVDs with the guidance of prior knowledge provided from the first stage. In summary, the main innovations in this study are listed as follows:

\begin{itemize}
  \item We proposed a two-stage framework to perform coarse-to-fine segmentation. The global spine correlation can be captured in the first stage, and the spatial information along the $z$ dimension can be exploited in the second stage. High computational memory results from the whole 3D spine MR images can be reduced using this two-stage framework.
  \item We incorporated a semi-supervised learning method, named CPS, into the proposed method in the first stage. CPS can alleviate the data imbalance problem, improve model generality, and provide global intra-slice information for the subsequent second stage.
  \item We introduced a novel attention-based module, referred to as CTAM, to fuse the crucial features related to intra- and inter-slice information in 3D MR images effectively. A comprehensive description of the spine can be achieved by using the CTAM. Hence, the challenges of anisotropic spatial resolution, inter-class similarity, and intra-class variety can be overcome, and the segmentation performance of VBs and IVDs can be improved.
  \item The proposed method was evaluated on a publicly available dataset containing 215 volumetric MR images with normal or abnormal spine structures. Moreover, the segmentation performance by the proposed method on the testing set, whose ground truth was publicly unavailable, is better than that of some state-of-the-art methods. Therefore, the proposed method may provide a new way for automatic and accurate segmentation of multiple structures with the data imbalance problem.
\end{itemize}

\section{Related Work}
\subsection{Spine Segmentation}
Traditional modeling methods, such as parametric model \cite{r6}, statistical shape model \cite{r1}, and data-driven geometric model \cite{r4}, were widely used in the previous studies for spine segmentation. However, the traditional modeling methods cannot accurately deal with the individual differences, thereby possibly limiting the segmentation performance \cite{r7}. Thus far, the spine segmentation methods based on deep learning have been widely researched \cite{r7, r8, r3, r9, r10}. Following the segmentation objects of the spine, these algorithms can be divided into sole segmentation of VBs or IVDs and simultaneous segmentation of multiple spinal structures. For the sole segmentation of VBs, InteractiveFCN \cite{r7} is a multi-task and one-stage algorithm. This algorithm labels and segments VB by referring to the prior knowledge of VB continuity. Tao et al. \cite{r8} proposed a two-stage Spine-Transformer method for VB segmentation. This method consists of a Transformer-based 3D object detector in the first stage to solve the labeling problem and a multi-task encoder-decoder network in the second stage for VB segmentation. Moreover, Zhang et al. \cite{r9} applied reinforcement learning to detect and segment the VBs, which can accumulate effective experience of the difference between VBs and background pixels. For the sole segmentation of IVDs, Li et al. \cite{r10} proposed an integrated multi-scale fully convolutional network with random modality voxel dropout learning for IVD localization and segmentation, which can improve the segmentation accuracy of IVDs. Although good segmentation performance is achieved, these methods \cite{r7, r8, r9, r10} focus on VB or IVD segmentation, which may lose some useful information on disease diagnosis of the spine.

Simultaneous segmentation of VBs and IVDs is more challenging than sole segmentation of VBs or IVDs because spatial relations of different structures should be considered in the former, and correctness of category labeling should be ensured. Spine-GAN \cite{r3} was designed to segment and classify six spine classes in one shot. It comprises a segmentation and a discriminative network, which can achieve high pixel segmentation accuracy. SpineParseNet \cite{r2} is a two-stage method that initially performs rough segmentation on 3D MR images and then conducts 2D refined segmentation. Although SpineParseNet can alleviate the problems of inter-class similarity and intra-class variation to some extent, it has some minor defects and disabilities in dealing with imbalanced data and blurry boundary issues \cite{r2}.

\subsection{Semi-Supervised Medical Image Segmentation}
In the field of medical image segmentation, labeled data are expensive and difficult to acquire, whereas the unlabeled data are abundant and easily available \cite{r11,r27}. Semi-supervised learning, which is proposed to take advantage of unlabeled and labeled data, can reduce the cost of data acquisition and improve the generalization ability of models. Thus, semi-supervised learning is a potential solution for medical image segmentation. Consistency regularization, one of the most popular semi-supervised learning methods, enforces the prediction of model consistency or similarity under the perturbation and tries to make the prediction decision boundary in a low-density regime \cite{r5,r28}. Moreover, consistency regularization is constructed based on smoothing hypothesis. It is helpful to prevent the overfitting of models and has been applied to medical image segmentation. For example, Li et al. \cite{r13} proposed an effective transformation-consistent self-ensembling model, which was trained with a Teacher–Student scheme and optimized by a weighted combination of supervised and unsupervised losses, for 2D and 3D medical image segmentation. Luo et al. \cite{r11} proposed a dual-task consistency-based framework to achieve the consistency of a pixel-level classification map and a level set representation via a task-transform layer for accurate medical image segmentation. Recently, Chen et al. \cite{r5} proposed a CPS method, which used consistency regularization with network perturbation. Unlike the Teacher–Student structure, CPS is combined with a self-training method that leverages the model’s confident prediction to generate pseudo labels for unlabeled data, thereby effectively avoiding the tight coupling between teacher and student networks \cite{dual_student}. Inspired by the idea proposed in \cite{r5}, we incorporated CPS into the spine segmentation task in this study to achieve stable segmentation performance and alleviate data imbalance to a certain extent.

\subsection{Attentional Feature Fusion}
Attention mechanism is widely used in medical image segmentation to improve the discriminative ability of features, reduce redundant information, and generate informative representations for accurate segmentation \cite{r17,r18,r19,r20,r21,r22}. Feature fusion, commonly used in image segmentation, can be applied to fuse multi-level \cite{r18}, multi-scale \cite{r19}, and multi-modality \cite{r20} features to obtain complementary information. Therefore, the combination of attention mechanism and feature fusion can take full advantage of these two strategies and further improve segmentation accuracy. Fan et al. \cite{r21} proposed a multi-scale attention network, which introduces a self-attention mechanism to integrate local features with their global dependencies adaptively for liver and tumor segmentation. Moreover, Li et al. \cite{r22} proposed a triple attention network to capture the global features of various dimensions, which can improve the feature discrimination ability and segmentation performance. However, only intra-slice information is captured, whereas inter-slice information associated with 3D volume is ignored in these methods \cite{r21, r22}. We proposed a novel attentional feature fusion method, named CTAM, to overcome this issue by effectively integrating the intra- and inter-slice features generated by 2D and 3D networks to achieve accurate spine segmentation.

\section{Method}
\begin{figure}
  \centering
  \includegraphics[width=8.8cm]{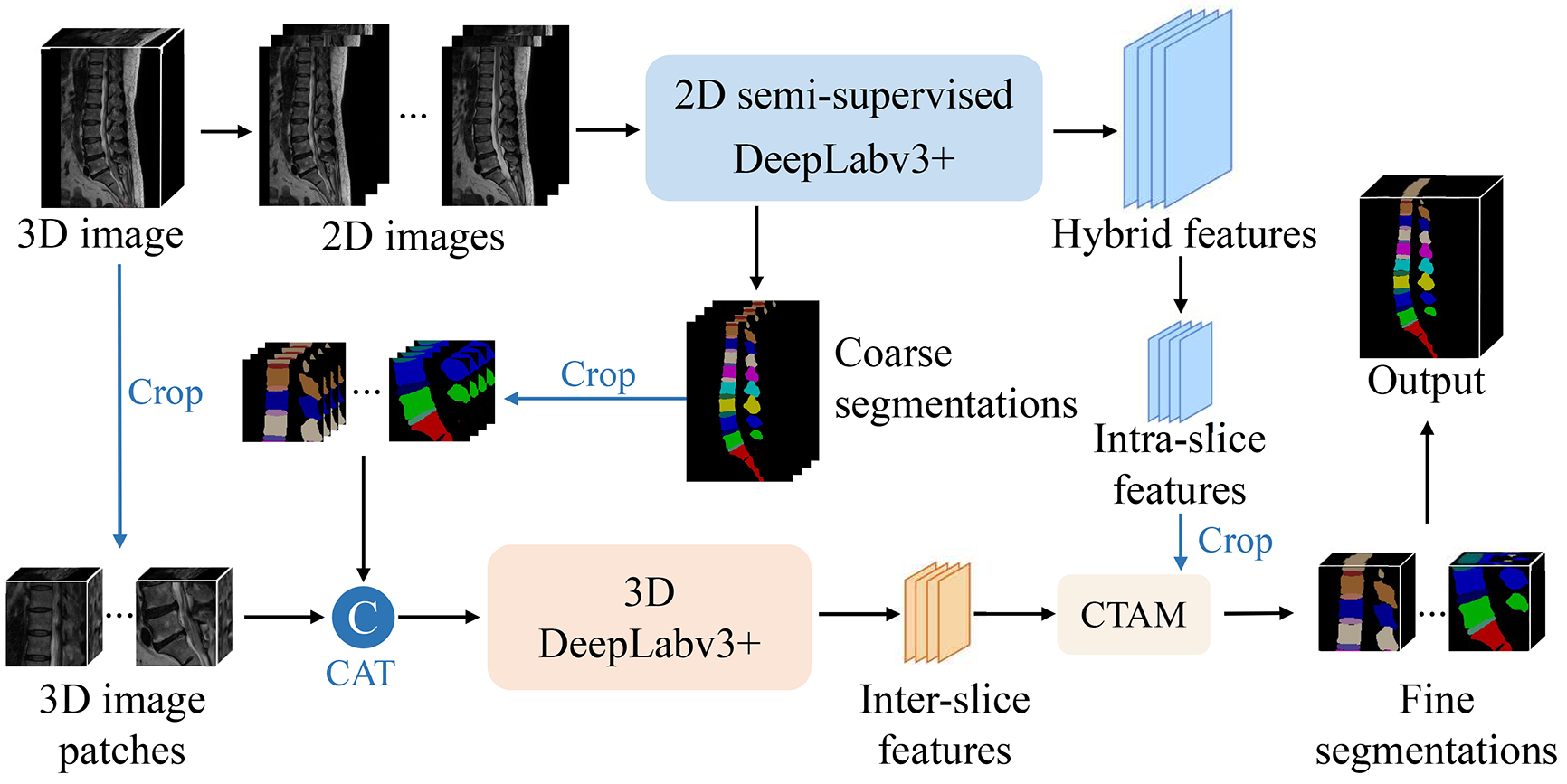}
  \caption{Architecture of SSHSNet, where “CAT” is short for concatenation.}
  \label{fig3}
\end{figure}

As illustrated in Fig. \ref{fig3}, the architecture of the proposed SSHSNet primarily consists of two stages. The first stage is constructed by two 2D DeepLabv3+ with semi-supervised learning strategy to exploit the rich intra-slice information, particularly the spatial correlations between the VBs and the IVDs. The coarse segmentation and the corresponding spatial correlations provided in this stage are used as prior knowledge for the next stage. The second stage consists of a 3D DeepLabv3+ and two 2D–3D fusion strategies to combine intra- and inter-slice information effectively for accurate spine segmentation. The details of the proposed method are provided in the following subsections.

\subsection{2D Semi-Supervised DeepLabv3+ for Coarse Segmentation}
In the first stage, our goal is to obtain spatial correlations of spinal structures in the downsampled 2D images for coarse segmentation. Hence, 2D DeepLabv3+ is used as the backbone, given its effectiveness in capturing contextual information at multiple scales \cite{r23}. Moreover, CPS is incorporated into the 2D network to use labeled and unlabeled data effectively, thereby improving the generalization ability of the proposed method.

\begin{figure}
  \centering
  \includegraphics[width=8.8cm]{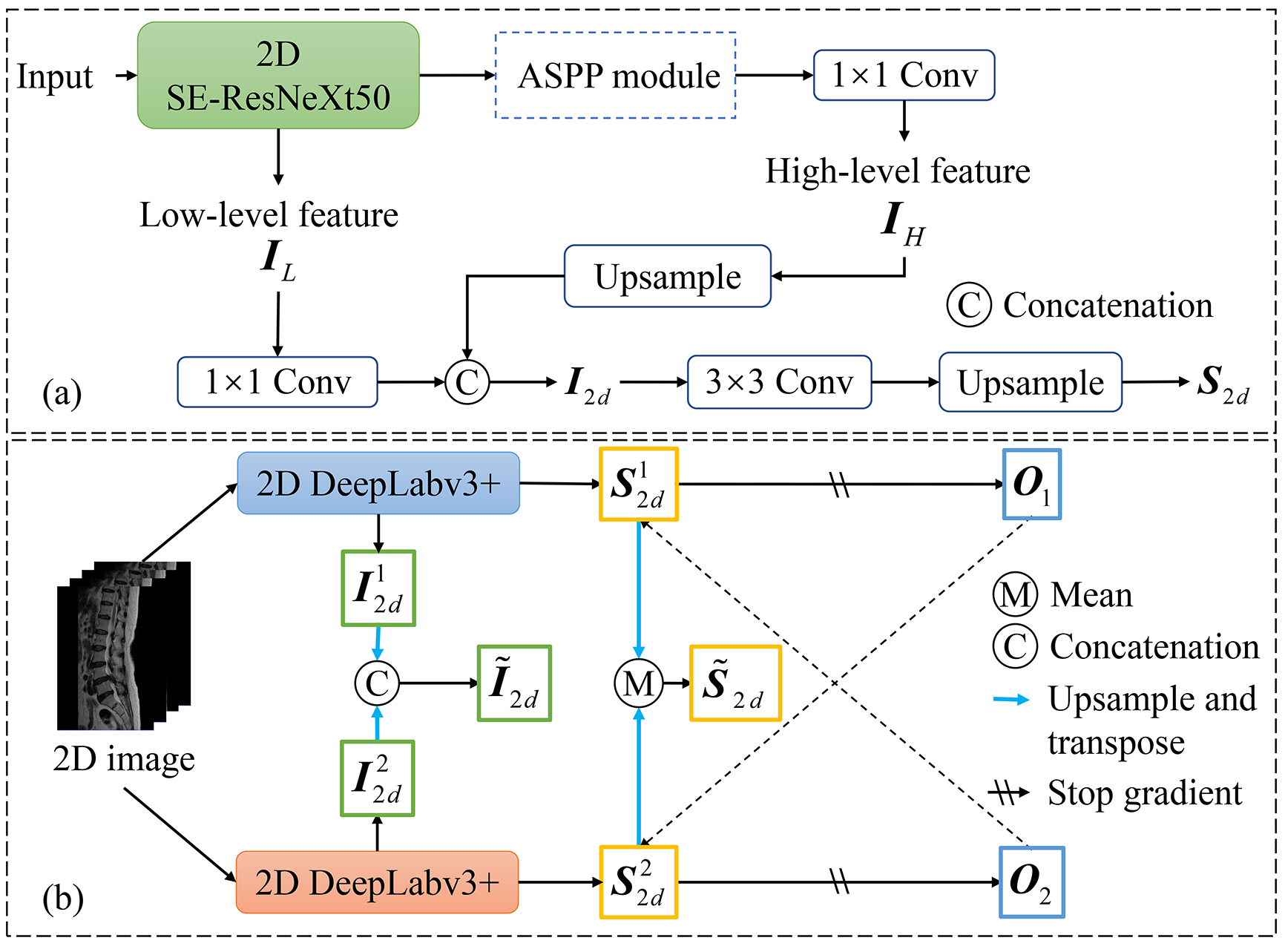}
  \caption{Structures of (a) 2D DeepLabv3+ and (b) CPS.}
  \label{fig2d}
\end{figure}

\subsubsection{2D DeepLabv3+ Network}

Given a set of 3D spine MR images $\boldsymbol{X}_{3 d} \in \mathbb{R}^{N \times Z \times W \times H}$, where $N$, $Z$, $W$, and $H$ are slice number, width, height, and subject number of 3D images, respectively. ${X}_{3 d}$ is initially split into 2D slices along $z$ dimension. Then, each 2D slice is combined with its adjacent two slices, and these three slices are downsampled once, which can be defined as $\boldsymbol{X}_{2 d} \in \mathbb{R}^{NZ \times \frac{W}{2} \times \frac{H}{2} \times 3}$. The 2D DeepLabv3+ used in this study follows the encoder–decoder structure (Fig. \ref{fig2d} (a)). The encoder is constructed by a SE-ResNext50 ($32 \times 4d$) \cite{senet, resnext} and an Atrous Spatial Pyramid Pooling (ASPP) module, whereas the decoder is built by twice upsampling with a factor of 4 to combine effectively the low- and high-level features provided by the encoder for a refined segmentation along object boundaries \cite{r23}. Moreover, SE-ResNext50 embraces the advantages of reduced network complexity and increased channel-wise correlations. With the well-trained 2D network, spatial and semantic information can be included in feature $\boldsymbol{I}_{2 d} \in \mathbb{R}^{NZ \times {C}_{2} \times \frac{W}{8} \times \frac{H}{8}}$, which is generated by combining low- and high-level features (Fig. \ref{fig2d} (a)). Moreover, the segmentation confidence map $\boldsymbol{S}_{2 d} \in \mathbb{R}^{NZ \times {C}_{1} \times \frac{W}{2} \times \frac{H}{2}}$ of 2D DeepLabv3+ can be achieved. This map contains probabilities of context pixels. Thus, $\boldsymbol{I}_{2 d}$ and $\boldsymbol{S}_{2 d}$ can be used as prior knowledge and are incorporated into the next stage to guide fine segmentation.
\begin{figure}
  \centering
  \includegraphics[width=8.8cm]{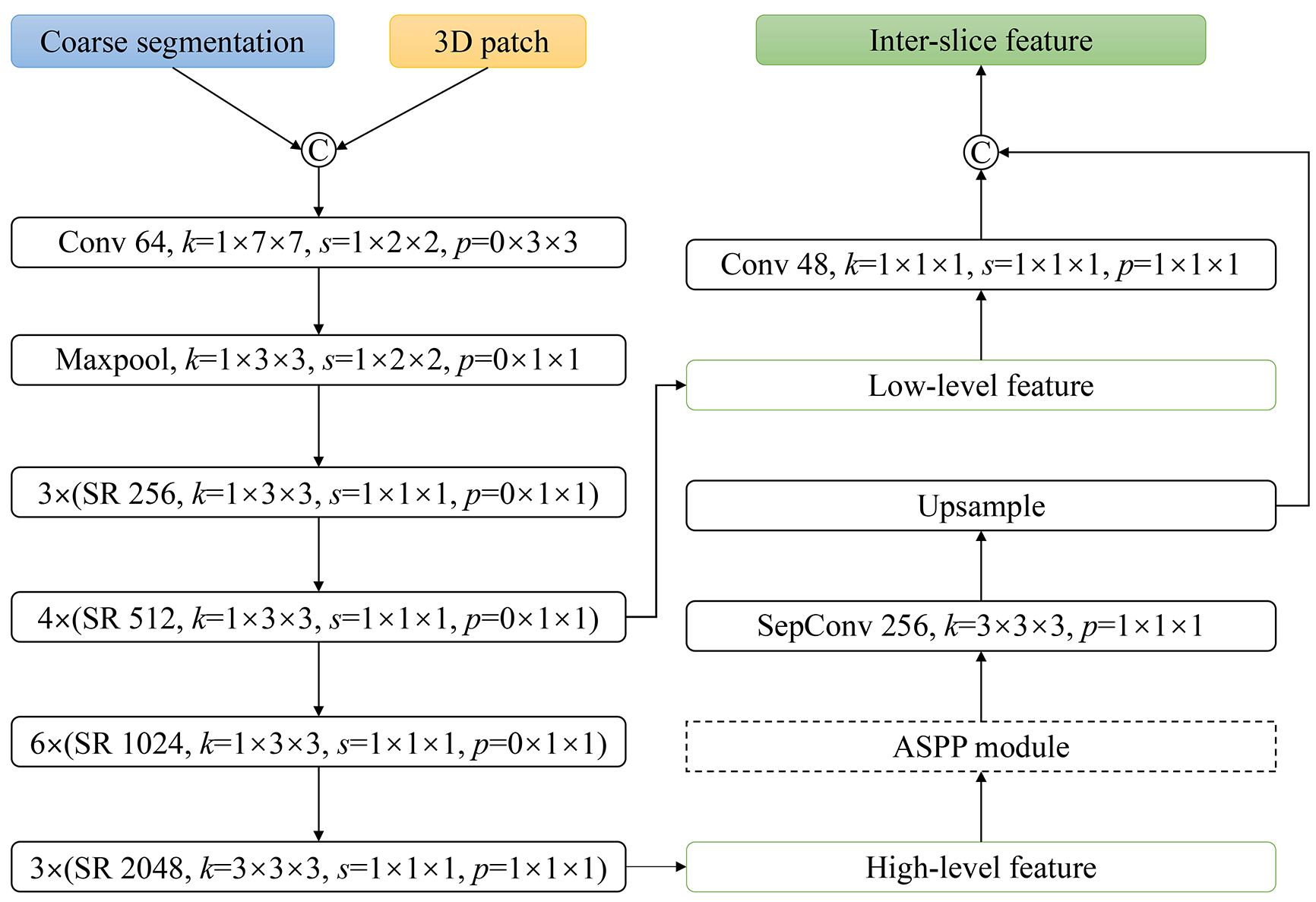}
  \caption{Structure of 3D DeepLabv3+, where “SR” and “SepConv” represent the SE-ResNext50 block and separable convolution, respectively. Moreover, “$k$”, “$s$”, and “$p$” represent kernel size, stride, and padding, respectively.}
  \label{fig4}
\end{figure}
\subsubsection{Cross Pseudo Supervision}
Although spatial and semantic information can be well captured by using the 2D DeepLabv3+ with SE-ResNext50 blocks, it is trained in a fully supervised way that is highly dependent on a large number of manual annotations. In spine segmentation, annotating sufficient training data with 19 spine structures is difficult. Moreover, the data with pathological changes, such as lumbar sacralization or lumbarization, are only a small proportion of the entire data, possibly resulting in decreased segmentation accuracy by using a fully supervised training method. Following the recent work \cite{r5}, we introduce an SSL method, named CPS (Fig. \ref{fig2d} (b)), by using self-training and consistency regularization with network perturbation to use unlabeled data effectively.

We first initialize two 2D DeepLabv3+ segmentation networks with different parameters $f_{2 d}\left(\boldsymbol{\theta}_{m}\right)$, where $m=1,2$. Labeled and unlabeled 2D data are downsampled once $\boldsymbol{X}_{2 d}^{\prime}=\left\{\boldsymbol{X}_{2 d}^{l}, \boldsymbol{X}_{2 d}^{u}\right\} \in \mathbb{R}^{NZ \times 3 \times \frac{W}{2} \times \frac{H}{2}}$ to be inputs of these networks. For the labeled data $\boldsymbol{X}_{2 d}^{l}$, ground truth can be used to supervise the network training, whereas the pseudo labels generated by a network are used to supervise the training of the other network for unlabeled data $\boldsymbol{X}_{2 d}^{u}$. In particular, let $\boldsymbol{S}_{2 d}^{m}=f_{2 d}\left(\boldsymbol{X}_{2 d}^{u} ; \boldsymbol{\theta}_{m}\right)$ represent the segmentation confidence map, where $\boldsymbol{S}_{2 d}^{m} \in \mathbb{R}^{NZ \times {C}_{1} \times \frac{W}{2} \times \frac{H}{2}}$. Moreover, $\boldsymbol{Y}_{2 d}^{m}=T(\boldsymbol{S}_{2 d}^{m})$ represents the pseudo label map generated from the corresponding segmentation confidence map $\boldsymbol{S}_{2 d}^{m}$, where $\boldsymbol{Y}_{2 d}^{m} \in \mathbb{R}^{NZ \times 1 \times \frac{W}{2} \times \frac{H}{2}}$. When training unlabeled data, the pseudo label $\boldsymbol{Y}_{2 d}^{1}$ supervises $f_{2 d}\left(\boldsymbol{\theta}_{2}\right)$ to prompt $\boldsymbol{Y}_{2 d}^{1}$ to have high similarity with $\boldsymbol{Y}_{2 d}^{1}$, and vice versa. A high similarity between the predictions of two perturbed networks for the same input can be achieved by using this training strategy. The unlabeled data with pseudo labels can be used as expanded training data. Therefore, good generality can be obtained by using the CPS method. Moreover, CutMix \cite{CutMix} is used to increase the training difficulty, enhance network stability, and achieve accurate segmentation. With the well-trained 2D network, we can achieve a hybrid feature map $\tilde{\boldsymbol{I}}_{2 d} \in \mathbb{R}^{N \times 2{C}_{3} \times Z \times \frac{W}{4} \times \frac{H}{4}}$ by fusing $\boldsymbol{I}_{2 d}^{1}$ and $\boldsymbol{I}_{2 d}^{2}$, and a coarse segmentation $\tilde{\boldsymbol{S}}_{2 d} \in \mathbb{R}^{N \times {C}_{1} \times Z \times W \times H}$ by fusing $\boldsymbol{S}_{2 d}^{1}$ and $\boldsymbol{S}_{2 d}^{2}$, which can be formulated as:
\begin{equation}
  \tilde{\boldsymbol{S}}_{2 d}=G\left(\boldsymbol{S}_{2 d}^{m}\right), \tilde{\boldsymbol{I}}_{2 d}=T\left(\boldsymbol{I}_{2 d}^{m}\right)
  \label{e1}
\end{equation}
where $G(\cdot)$ represents the operations of upsampling, transposing, and average. Moreover, $T(\cdot)$ represents the operations of upsampling, transposing, and concatenation.

\subsection{3D Full-Resolution Patch-Based DeepLabv3+ for Fine Segmentation}
In the first stage, the 2D semi-supervised DeepLabv3+ network can provide global intra-slice information ($\tilde{\boldsymbol{I}}_{2 d}$ and $\tilde{\boldsymbol{S}}_{2 d}$) that contains relationships among adjacent spine connections. However, it neglects the inter-slice information containing spatial information along the $z$ dimension. Moreover, the detailed information is lost because downsampling is performed on 2D images. Additionally, 3D networks with whole 3D images as inputs entail high computational costs. In the second stage, we propose a 3D full-resolution patch-based DeepLabv3+ (Fig. \ref{fig4}) together with two 2D–3D fusion strategies to address the issues mentioned by effectively integrating the intra- and inter-slice information for accurate spine segmentation.

Given the high computational cost in the 3D network, CPS is excluded from the 3D DeepLabv3+. The feature space should be first aligned to fuse the features generated from the 2D and 3D networks. Therefore, the hybrid feature $\tilde{\boldsymbol{I}}_{2 d}$ and coarse segmentation $\tilde{\boldsymbol{S}}_{2 d}$ from the 2D semi-supervised DeepLabv3+ are transformed to 3D patches as follows:
\begin{equation}
  \tilde{\boldsymbol{S}}_{3 d}, \tilde{\boldsymbol{I}}_{3 d}=\phi\left(\tilde{\boldsymbol{S}}_{2 d}, \tilde{\boldsymbol{I}}_{2 d}\right)
  \label{e2}
\end{equation}
where $\tilde{\boldsymbol{I}}_{3 d} \in \mathbb{R}^{NB \times 2 {C}_{3} \times Z \times \frac{w}{4} \times \frac{h}{4}}$ and $\tilde{\boldsymbol{S}}_{3 d} \in \mathbb{R}^{NB \times {C}_{1} \times Z \times w \times h}$ represent the 3D intra-slice feature and coarse segmentation, respectively. $B$ is 3D patch number of each sample. $w$ and $h$ are the weight and height of the 3D image patch, respectively. $\phi(\cdot)$ represents the operations of category-balanced cropping and concatenation. Moreover, a convolution layer with kernel size $1 \times 1\times 1$ is applied on $\tilde{\boldsymbol{I}}_{3 d}$ to reduce its channel number, and thus, a new 3D intra-slice feature $\tilde{\boldsymbol{I}}^{c}_{3 d} \in \mathbb{R}^{NB \times {C}_{3} \times Z \times \frac{w}{4} \times \frac{h}{4}}$ can be achieved. Subsequently, two different fusion strategies are applied in the proposed method. First, the 3D patches $\boldsymbol{X}_{3 d}^{p} \in \mathbb{R}^{NB \times 1 \times Z \times w \times h}$ cropped from volumetric images are concatenated with the coarse segmentation $\tilde{\boldsymbol{S}}_{3 d}$ to be inputs $\tilde{\boldsymbol{M}}_{1} \in \mathbb{R}^{NB \times \left(1+C_{1}\right) \times Z \times w \times h}$ of 3D DeepLabv3+. Therefore, the training of 3D DeepLabv3+ is based not only on the features probed from the original image patches but also on the probabilities of the context pixels from the 2D network. The learning efficiency of the 3D network can be improved with the guidance from the probabilities of context pixels. Similar to 2D DeepLabv3+, a feature map $\tilde{\boldsymbol{I}}_{2 d-3 d} \in \mathbb{R}^{NB \times {C}_{3} \times Z \times \frac{w}{4} \times \frac{h}{4}}$ combining low- and high-level features can be obtained:
\begin{equation}
  \tilde{\boldsymbol{I}}_{2 d-3 d}=V_{3 d}\left(\tilde{\boldsymbol{M}}_{1}\right)
  \label{e3}
\end{equation}
where $V_{3 d}\left( \cdot \right)$ represents the function of 3D DeepLabv3+.

In the second fusion strategy, a CTAM is applied to fuse $\tilde{\boldsymbol{I}}^{c}_{3 d}$ and $\tilde{\boldsymbol{I}}_{2 d-3 d}$ to combine intra- and inter-slice features effectively. This approach can be described as Eq. \ref{e4}. The detail of CTAM is introduced in the next section. The attention mechanism is usually performed on the feature level. Thus, the CTAM is excluded from the fusion between the 3D image patch $\boldsymbol{X}_{3 d}^{p}$ and coarse segmentation $\tilde{\boldsymbol{S}}_{3 d}$.
\begin{equation}
  \tilde{\boldsymbol{M}}_{2}=\xi\left(\tilde{\boldsymbol{I}}^{c}_{3 d}, \tilde{\boldsymbol{I}}_{2 d-3 d}\right)
  \label{e4}
\end{equation}
where $\xi(\cdot)$ represents the function of CTAM and $\tilde{\boldsymbol{M}}_{2} \in \mathbb{R}^{NB \times C_{4} \times  Z \times \frac{w}{4} \times \frac{h}{4}}$ represents the cross-fused feature generated from CTAM. The fine segmentation provided by the proposed method can be defined as follows:
\begin{equation}
  \boldsymbol{Y}_{3 d}^{p}=\psi\left(\tilde{\boldsymbol{M}}_{2}\right)
  \label{e5}
\end{equation}
where $\psi(\cdot)$ represents the prediction operation of voxel probability in 3D network and $\boldsymbol{Y}_{3d}^{p} \in \mathbb{R}^{NB \times C_{1} \times Z \times \frac{w}{4} \times \frac{h}{4}}$ represents the fine segmentation of the 3D network. Finally, spine segmentation of whole 3D MR images $\boldsymbol{Y}_{3d}$ can be restored from the patch-based results $\boldsymbol{Y}_{3d}^{p}$.

\subsection{The Cross Tri-Attention Module}
\begin{figure}
  \centering
  \includegraphics[width=8.8cm]{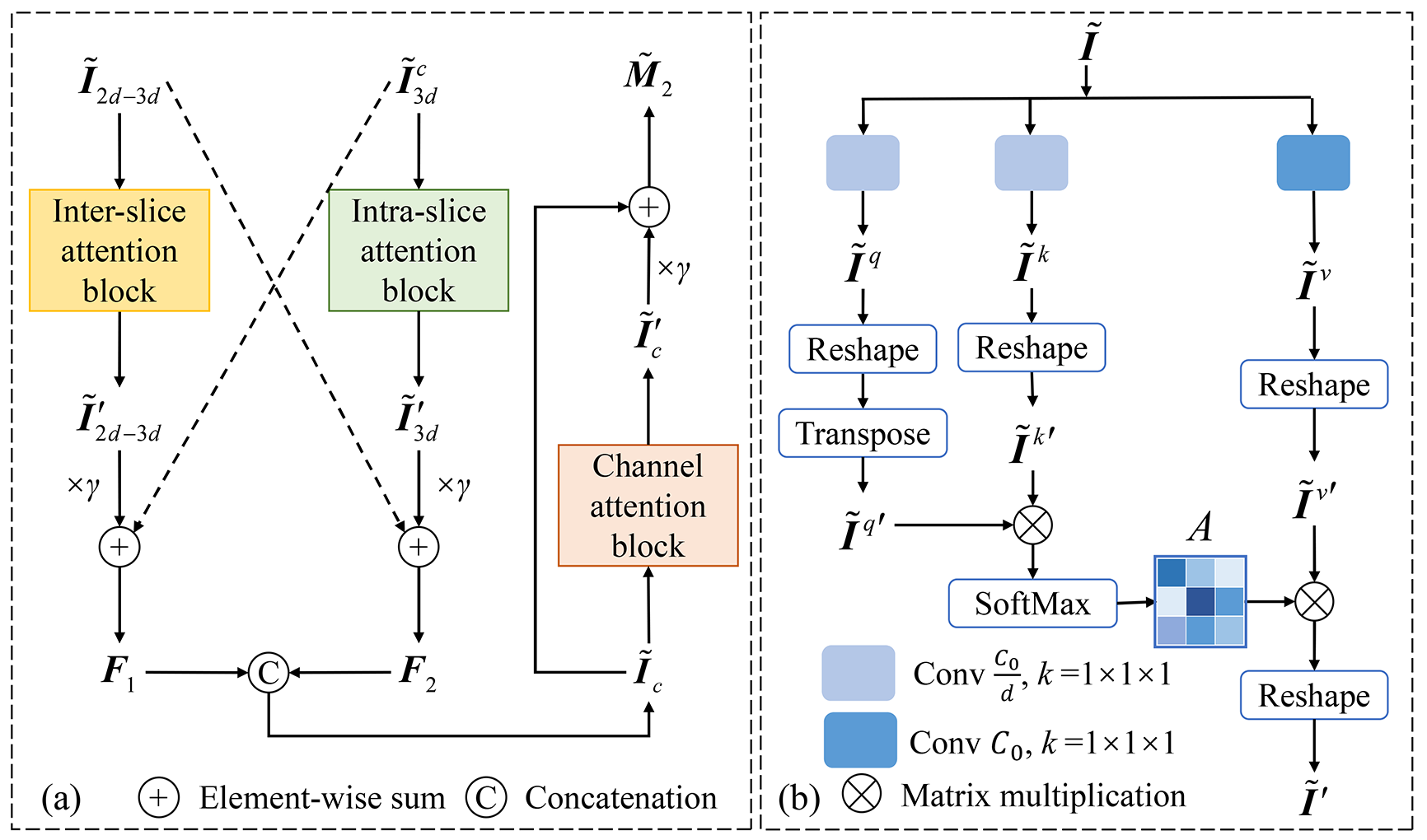}
  \caption{(a) Architecture of the proposed CTAM. (b) Operation details of an attention block. “Conv $t$” denotes a convolution with $t$ output channels, and “$k$” represents the kernel size. Moreover, $d$ equals to 8, 8, and 1 in inter-slice, intra-slice, and channel attention block, respectively.}
  \label{fig5}
\end{figure}

Fusing intra- and inter-slice features is crucial because their complementary information can provide a comprehensive description of the spine and improve segmentation performance. However, redundancy is prone to occur if simple concatenation is conducted on intra- and inter-slice features. We propose a novel feature fusion based on attention mechanism, which is referred to as CTAM, to alleviate the problem mentioned. As shown in Fig. \ref{fig5} (a), a CTAM comprises three attention blocks: inter-slice, intra-slice, and channel attention blocks. The operations for each attention block are similar except for the size of feature maps generated from each operation.

We assumed that the feature $\tilde{\boldsymbol{I}}$ is the input of each attention block (Fig. \ref{fig5} (b)). $\tilde{\boldsymbol{I}}$ is initially fed into three convolution layers to generate three feature maps $\tilde{\boldsymbol{I}}^{q} \in \mathbb{R}^{NB \times \frac{C_{0}}{d} \times Z \times \frac{w}{4} \times \frac{h}{4}}$, $\tilde{\boldsymbol{I}}^{k} \in \mathbb{R}^{NB \times \frac{C_{0}}{d} \times Z \times \frac{w}{4} \times \frac{h}{4}}$, and $\tilde{\boldsymbol{I}}^{v} \in \mathbb{R}^{NB \times C_{0} \times Z \times \frac{w}{4} \times \frac{h}{4}}$, where, $C_{0}$ represents the channel of the feature $\tilde{\boldsymbol{I}}$. In particular, $C_{0}$ and $d$ equal to $C_{3}$ and 8, respectively, in inter- and intra-slice attention blocks. Moreover, $C_{0}$ and $d$ equal to $ C_{4}$ and 1, respectively, in channel attention block. Then, the $\tilde{\boldsymbol{I}}^{k}$ and $\tilde{\boldsymbol{I}}^{v}$ are reshaped into two feature maps $\tilde{\boldsymbol{I}}^{k \prime} \in \mathbb{R}^{NB \times \alpha \times \frac{\beta}{d}}$ and $\tilde{\boldsymbol{I}}^{v \prime} \in \mathbb{R}^{NB \times \alpha \times \beta}$, respectively, where $\alpha$ is $Z$, $\frac{wh}{16}$, and $C_{0}$ in inter-slice, intra-slice and channel attention block, respectively, and $\beta$ is $\frac{C_{0}w h}{16}$, $ C_{0} Z$, and $\frac{Zwh}{16}$ in inter-slice, intra-slice, and channel attention blocks, respectively. Moreover, $\tilde{\boldsymbol{I}}^{q}$ is reshaped and transposed into a feature map $\tilde{\boldsymbol{I}}^{q \prime} \in \mathbb{R}^{NB \times \frac{\beta}{d} \times \alpha}$. Subsequently, matrix multiplication between the $\tilde{\boldsymbol{I}}^{k \prime}$ and $\tilde{\boldsymbol{I}}^{q \prime}$ is performed, and a SoftMax layer is applied after the multiplication to calculate the attention weight $\boldsymbol{A} \in \mathbb{R}^{\alpha \times \alpha}$. Then, a matrix multiplication between $\tilde{\boldsymbol{I}}^{v \prime}$ and $\boldsymbol{A}$ is conducted. Finally, the multiplication result is reshaped, and a new attention feature $\tilde{\boldsymbol{I}}^{\prime} \in \mathbb{R}^{NB \times C_{0} \times Z \times \frac{w}{4} \times \frac{h}{4}}$ can be obtained. Therefore, inter-slice feature $\tilde{\boldsymbol{I}}^{\prime}_{2 d-3 d} \in \mathbb{R}^{NB \times C_{3} \times Z \times \frac{w}{4} \times \frac{h}{4}}$ and intra-slice feature $\tilde{\boldsymbol{I}}_{3 d}^{\prime} \in \mathbb{R}^{NB \times C_{3} \times Z \times \frac{w}{4} \times \frac{h}{4}}$ can be achieved by using intra- and inter-slice attention blocks. We first multiply $\tilde{\boldsymbol{I}}_{2 d-3 d}^{\prime}$ by a scale parameter $\gamma$ to fuse intra-slice and inter-slice information effectively and then perform an element-wise sum operation with $\tilde{\boldsymbol{I}}_{3 d}$ to obtain $\boldsymbol{F}_{1} \in \mathbb{R}^{NB \times C_{3} \times Z \times \frac{w}{4} \times \frac{h}{4}}$. Similarly, $\boldsymbol{F}_{2} \in \mathbb{R}^{NB \times C_{3} \times Z \times \frac{w}{4} \times \frac{h}{4}}$ can be achieved via the fusion of $\tilde{\boldsymbol{I}}_{3 d}^{\prime}$ and $\tilde{\boldsymbol{I}}_{2 d-3 d}$. Then, $\boldsymbol{F}_{1}$ and $\boldsymbol{F}_{2}$ are concatenated to be $\tilde{\boldsymbol{I}}_{c} \in \mathbb{R}^{NB \times C_{4} \times Z \times \frac{w}{4} \times \frac{h}{4}}$. Finally, the channel attention block is performed on $\tilde{\boldsymbol{I}}_{c}$ to filter the redundant information among the channel relationship and obtain the final cross-fused feature $\tilde{\boldsymbol{M}}_{2}$.

\subsection{Loss Function}
The training objective for the 2D semi-supervised DeepLabv3+ contains two losses (listed in Eq. \ref{e6}): supervision loss $L_{s}$ and CPS loss $L_{cps}$.
\begin{equation}
  L_{2 d}=L_{s}+\lambda L_{c p s}
  \label{e6}
\end{equation}
where $\lambda$ is the trade-off weight. In particular, the supervision loss on labeled data consists of cross-entropy and dice losses, which can be defined as follows:
\begin{equation}
  L_{s}=\sum_{m=1}^{2}\left(L_{c e}\left(\boldsymbol{Y}_{m}^{*}, \boldsymbol{S}_{2 d}^{m} \right)+ L_{d c}\left(\boldsymbol{Y}_{m}^{*}, \boldsymbol{S}_{2 d}^{m} \right)\right)
  \label{e7}
\end{equation}
where $\boldsymbol{Y}^{*}$ represents the ground truth. Moreover, CPS loss on unlabeled data can be formulated as follows:
\begin{equation}
  L_{c p s}=L_{c e}\left(\boldsymbol{Y}_{2 d}^{1}, \boldsymbol{S}_{2 d}^{2} \right)+L_{c e}\left(\boldsymbol{Y}_{2d}^{2}, \boldsymbol{S}_{2 d}^{1} \right)
  \label{e8}
\end{equation}
For the training objective of 3D full-resolution patch-based DeepLabv3+, the loss criterion $L_{3 d}$ is the linear combination of cross-entropy and dice losses:
\begin{equation}
  L_{3 d}=L_{c e}\left(\boldsymbol{Y}_{3 d}^{*}, \boldsymbol{Y}_{3 d} \right)+L_{d c}\left(\boldsymbol{Y}_{3 d}^{*}, \boldsymbol{Y}_{3 d} \right)
  \label{e9}
\end{equation}
Moreover, the cross-entropy and dice losses can be formulated as follows:
\begin{equation}
  L_{c e}\left(\boldsymbol{O}^{*},\boldsymbol{R} \right)=-\frac{1}{n \mu} \sum_{i=1}^{n \mu} \boldsymbol{o}_{i}^{*} \log \boldsymbol{r}_{i}
  \label{e10}
\end{equation}
\begin{equation}
  L_{d c}\left(\boldsymbol{O}^{*},\boldsymbol{R} \right)=-\frac{1}{n} \frac{2 \sum_{i=1}^{\mu} \boldsymbol{o}_{i}^{*} \boldsymbol{r}_{i}}{\sum_{i=1}^{\mu} \left(\boldsymbol{o}_{i}^{*} + \boldsymbol{r}_{i}\right)}
  \label{e11}
\end{equation}
where $\left\{\boldsymbol{o}_{i}^{*}\right\} \in \boldsymbol{O}^{*}$ represents the pixel of ground truth. Moreover, $\left\{\boldsymbol{r}_{i}\right\} \in \boldsymbol{R}$ represents the pixel of segmentation result. $n$ equals to $N_{l} Z$, $N_{u} Z$, and $N_{l} B$ in Eqs. \ref{e7}, \ref{e8}, and \ref{e9}, respectively. $N_{l}$ and $N_{u}$ represent the number of labeled and unlabeled data, respectively. Moreover, $\mu$ equals to $\frac{W H}{4}$, $\frac{W H}{4}$, and $Z w h$ in Eqs. \ref{e7}, \ref{e8}, and \ref{e9}, respectively.

\section{Experiment Setup}
\subsection{Materials}
A total of 215 T2-weighted MR volumetric images of the spine are provided by the China Society of Image and Graphics Challenge on Automated Multi-class Segmentation of Spinal Structures on Volumetric MR Images (MRSpineSeg Challenge, https://www.spinesegmentation-challenge.com), where 172 images are divided as training data with publicly available segmentation masks and 43 images are set as testing data with publicly unavailable segmentation masks. The in-plane resolution of the MR images ranges from 0.30 mm to 0.59 mm, and the slice spacing ranges from 4.40 mm to 5.50 mm. The in-plane sizes range from $880 \times 880$ to $990 \times 990$, and the slice number ranges from 12 to 18. All images include thoracic, lumbar, and sacral vertebrae, in which experts marked and segmented 10 types of VBs and 9 types of IVDs (Fig. 1). However, not all images contain 19 vertebral structures because some of the corresponding patients suffered from spinal disorders. The detailed information about the spinal disorders in this dataset can be referred to \cite{r2}.

\subsection{Preprocessing}
All 3D spine MR images were preprocessed as follows. First, the median resolution was counted from the training set, and all images were resampled to the median resolution $0.34 \times 0.34 \times 4.4$ $mm^{3}$. Second, the zero background was removed to reduce the unnecessary computational costs. Third, image intensities were normalized using z-score method. Finally, all images were cropped and used as inputs of the proposed method.

\subsection{Implementation Details}
Nine spine segmentation methods, including four state-of-the-art methods (i.e., 2D and 3D nnUnet \cite{r24}, CoTr \cite{r25}, and SpineParseNet \cite{r2}), the proposed method, and five variations of the proposed method were used to assess the segmentation performance. The configurations of different methods are as follows:
\begin{itemize}
  \item nnUnet: It is a Unet-based segmentation method that automatically configures itself. These self-configuration processes include preprocessing, network architecture, training, and post-processing. Moreover, 2D and 3D networks were implemented. For 2D and 3D networks, the patches with sizes of $512 \times 896$ and $320 \times 640 \times 8$ were used as inputs, respectively. Moreover, both the 2D and 3D networks were trained for 1000 epochs.
  \item CoTr: Self-attention mechanism in the Transformer is incorporated into the 3D convolutional neural network (CNN). Only a small set of key positions is applied in the Transformer, which can be used to process the multi-scale feature maps produced by the CNN and maintain abundant high-resolution information for segmentation. Moreover, the code of CoTr, which is based on nnUnet, is publicly available. Therefore, the implemented details of CoTr are the same as those of the 3D network used in nnUnet, except for the network architecture.
  \item SpineParseNet: This two-stage framework consists of a 3D graph convolutional segmentation network for 3D coarse segmentation and a 2D residual U-Net for 2D segmentation refinement. The testing data used in SpineParseNet \cite{r2} are the same as those used in this study. Thus, the results of the testing data presented in \cite{r2} are directly used for comparison in this study.
  \item SSHSNet-2D-w/o-Pre-CPS: It comprises the proposed 2D DeepLabv3+ network with randomly initialized weights.
  \item SSHSNet-2D-w/o-CPS: The proposed 2D DeepLabv3+ network is trained based on a pretrained model.
  \item SSHSNet-2D-CPS: The semi-supervised 2D DeepLabv3+ network used in the proposed method.
  \item SSHSNet-3D: It comprises the proposed 3D DeepLabv3+ network with randomly initialized weights.
  \item SSHSNet-w/o-CTAM: CTAM is ignored, and intra- and inter-slice features are concatenated directly in the proposed method.
\end{itemize}

The training details for the proposed method and its variations are as follows. The input for SSHSNet-2D-w/o-Pre-CPS, SSHSNet-2D-w/o-CPS, and SSHSNet-2D-CPS was downsampled once, with a size of $3 \times W \times H$, where $W$ and $H$ were set to 224 and 440, respectively. The epoch number and batch size were set to 1000 and 4, respectively. Moreover, the learning rate was set to 0.001 initially and was lowered by 10 times at epoch 50 and 400. Transfer learning with a pretrained 2D DeepLabv3+ network, which was trained on ImageNet, was applied for SSHSNet-2D-w/o-CPS and SSHSNet-2D-CPS. Moreover, testing data were used as unlabeled data during the training of SSHSNet-2D-CPS because their segmentation masks were unavailable. For 3D network used in the proposed method, the input was with size of $Z \times w \times h$, where $Z$, $w$, and $h$ were set to 12, 192, and 192, respectively. The epoch number and batch size were set to 150 and 4, respectively. The learning rate was set to 0.001 initially and was lowered by 10 times at epoch 25 and 100. Moreover, parameter $\lambda$ in Eq. \ref{e6} were set to 1 empirically. However, instead of using 3D full-resolution patch (i.e., $Z \times w \times h$), 3D volumes down-sampled once with size of $224 \times 440 \times 12$ were used as inputs of the SSHSNet-3D method to include spatial information for fair comparison. The epoch number and batch size were set to 1000 and 4, respectively. Additionally, the learning rate was set to 0.001 initially and was lowered by 10 times at epoch 50 and 400. Moreover, we set up $C_{1}$, $C_{2}$, $C_{3}$ and $C_{4}$ to 20, 128, 256 and 512, respectively, where $C_{1}$ is category number (i.e., 19 spine structures and background).

The proposed method was implemented by using PyTorch, and all experiments were performed on a server with one NVIDIA GeForce 2080Ti GPU.
\subsection{Evaluation Method}
In the experiments, a fivefold cross-validation method was initially performed on the training data to evaluate the effectiveness of the proposed method and its variations. Five segmentation models and their corresponding results can be achieved with the fivefold cross-validation method. Thus, the segmentation result on training data was reported as the average value of the fivefold results. All testing data were initially fed into the five segmentation models trained on the training data to achieve segmentation results of testing data. Then, five segmentation results can be provided by the five segmentation models. Finally, the average of the five segmentation results can be achieved and uploaded to the public evaluation website (https://www.spinesegmentation-challenge.com/) to assess model performance on the testing data.

Dice similarity coefficient (DSC) was used as the quantitative metric to assess the segmentation performance. In particular, mean DSC was calculated for an individual spine structure (i.e., VB or IVD) in the original image space and averaged across all subjects. Moreover, the overall mean DSC was calculated by using the average value of the fivefold results.

\section{Experimental Results}
\subsection{Comparison of 2D and 3D DeepLabv3+}
\begin{figure}
  \centering
  \includegraphics[width=8.8cm]{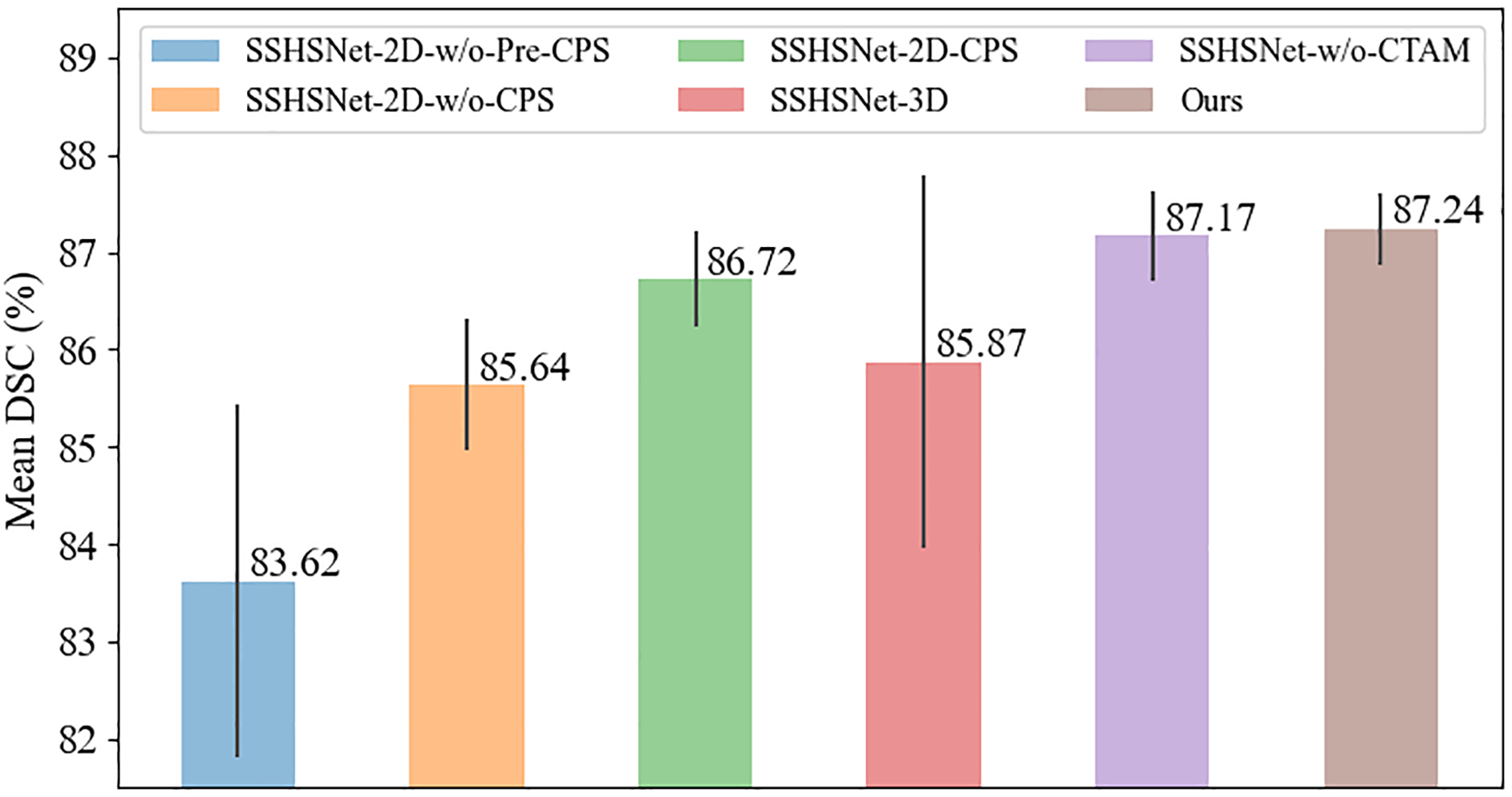}
  \caption{Overall mean DSC (\%) of the proposed method and its variations on the training set by using fivefold cross validation.}
  \label{fig6}
\end{figure}

In this section, comparisons were conducted among SSHSNet-2D-w/o-Pre-CPS, SSHSNet-2D-w/o-CPS, and SSHSNet-3D. As expected, the mean DSC value of SSHSNet-3D are higher than those of SSHSNet-2D-w/o-Pre-CPS with the same in-plane resolution as inputs and the same condition of the initialized parameter (Fig. \ref{fig6}), which benefits from the inter-slice information included in SSHSNet-3D. The segmentation performance achieved by using SSHSNet-2D-w/o-CPS is higher than that achieved by using SSHSNet-2D-w/o-Pre-CPS (Fig. \ref{fig6} and Table \ref{t1}). The overall mean DSC in SSHSNet-2D-w/o-CPS is slightly lower than that in SSHSNet-3D (Fig. \ref{fig6}). However, compared with SSHSNet-3D, SSHSNet-2D-w/o-CPS realizes the coveted segmentation performance in most VB and IVD segmentations (Table \ref{t1}). This finding indicates that the training based on a pretrained model plays an important role in achieving promising results for the network. Similar observations can be determined in the visualization results in the 3rd, 4th, and 6th columns in Fig. \ref{fig7}.

\subsection{Effectiveness of Two Stage Combination}
Semi-supervised 2D (SSHSNet-2D-CPS) and 3D (SSHSNet-3D) networks were performed separately to evaluate the effectiveness of the proposed two-stage framework. Given that 2D–3D feature fusion is unavailable in the SSHSNet-2D-CPS or SSHSNet-3D methods, the CTAM was replaced by a simple feature concatenation in the proposed method, i.e., SSHSNet-w/o-CTAM, to make a fair comparison. As shown in Table \ref{t1}, the segmentation performance of SSHSNet-2D-CPS is better than that of SSHSNet-3D, which may benefit from the semi-supervised learning used in the 2D network. Moreover, the segmentation accuracy obtained by using SSHSNet-w/o-CTAM is higher than that obtained by using SSHSNet-2D-CPS or SSHSNet-3D in most VB and IVD segmentations. This observation verifies the effectiveness of the two-stage strategy based on rough and fine segmentation. The visualization results in the 5th to 7th columns in Fig. \ref{fig7} show that compared with SSHSNet-2D-CPS and SSHSNet-3D, SSHSNet-w/o-CTAM displays superior segmentation performance. This finding indicates that the two-stage algorithm is more suitable for accurate spine segmentation than the one-stage algorithms.

	\begin{landscape}
\renewcommand\arraystretch{1.2}
\setlength\tabcolsep{3pt}
\begin{table*}
  \scriptsize
  \centering
  \caption{Mean DSC (\%) of the proposed method and its variations for individual VB and IVD segmentation on the training set by using fivefold cross validation.}
  \label{t1}
  \begin{tabular}{llllllllllllllllllll}
\hline
                       & S              & L5             & L4             & L3             & L2             & L1             & T12            & T11            & T10            & T9             & L5/S           & L4/L5          & L3/L4          & L2/L3          & L1/L2          & T12/L1         & T11/T12        & T10/T11        & T9/T10         \\ \hline
SSHSNet-2D-w/o-Pre-CPS & 82.58          & 84.83          & 85.46          & 85.68          & 85.13          & 85.38          & 83.38          & 76.14          & 51.99          & 0.00           & 84.17          & 84.36          & 87.31          & 86.88          & 87.39          & 86.63          & 81.73          & 65.32          & 0.00           \\
SSHSNet-2D-w/o-CPS     & 85.72          & 87.20          & 87.73          & 87.78          & 87.55          & 87.80          & 85.87          & 81.17          & 62.10          & 0.00           & 86.24          & 86.33          & 88.86          & 88.21          & 88.75          & 88.20          & 84.97          & 71.91          & 0.00           \\
SSHSNet-2D-CPS         & 87.30          & 88.09          & 88.72          & 88.99          & 88.80          & 88.83          & 86.62          & 82.17          & 57.27          & 61.30          & 87.12          & 86.60          & 89.62          & 89.20          & 89.43          & 89.09          & 84.34          & 75.69          & 57.18          \\
SSHSNet-3D             & 86.17          & 86.84          & 87.39          & 87.70          & 87.31          & 87.63          & 85.87          & 81.15          & 56.34          & 0.00           & 85.67          & 85.60          & 88.40          & 87.93          & 88.26          & 87.35          & 83.14          & 65.87          & 0.00           \\
SSHSNet-w/o-CTAM       & \textbf{87.96} & \textbf{88.67} & 89.12          & 89.32          & 89.10          & 89.04          & 87.05          & \textbf{83.82} & 60.35          & 55.83          & 87.24          & 86.88          & \textbf{90.02} & 89.47          & 89.38          & 89.06          & 85.64          & \textbf{77.06} & 56.66          \\
Ours                   & 87.91          & 88.66          & \textbf{89.60} & \textbf{89.56} & \textbf{89.64} & \textbf{90.20} & \textbf{89.53} & 83.23          & \textbf{71.72} & \textbf{75.77} & \textbf{87.71} & \textbf{88.31} & 89.95          & \textbf{89.88} & \textbf{90.76} & \textbf{90.41} & \textbf{88.42} & 70.02          & \textbf{84.36}\\ \hline
\end{tabular}
\end{table*}
		\end{landscape}

\subsection{Effectiveness of Semi-supervised Learning}
CPS was removed from the 2D network to investigate the influence of semi-supervised learning on the performance of the proposed method, and comparison was performed on the proposed 2D network with/without CPS, i.e., SSHSNet-2D-w/o-CPS and SSHSNet-2D-CPS. As shown in Table \ref{t1}, the segmentation accuracy of the SSHSNet-2D-CPS is higher than that of the SSHSNet-2D-w/o-CPS. In particular, SSHSNet-2D-w/o-CPS fails to segment some spine structures with unbalanced data labels, such as T9 VB and T9/T10 IVD, whereas SSHSNet-2D-CPS can segment these spine structures to some extent. This finding indicates that model generalization ability can be improved by incorporating CPS into the proposed method. Moreover, the 4th and 5th columns show that T9 VB and T9/T10 IVD can be well segmented by using SSHSNet-2D-CPS, whereas these two spine structures are abandoned in SSHSNet-2D-w/o-CPS. This observation further proves the effectiveness of incorporating CPS into the proposed method.

\subsection{Effectiveness of CTAM}
\begin{figure}
  \centering
  \includegraphics[width=8.8cm]{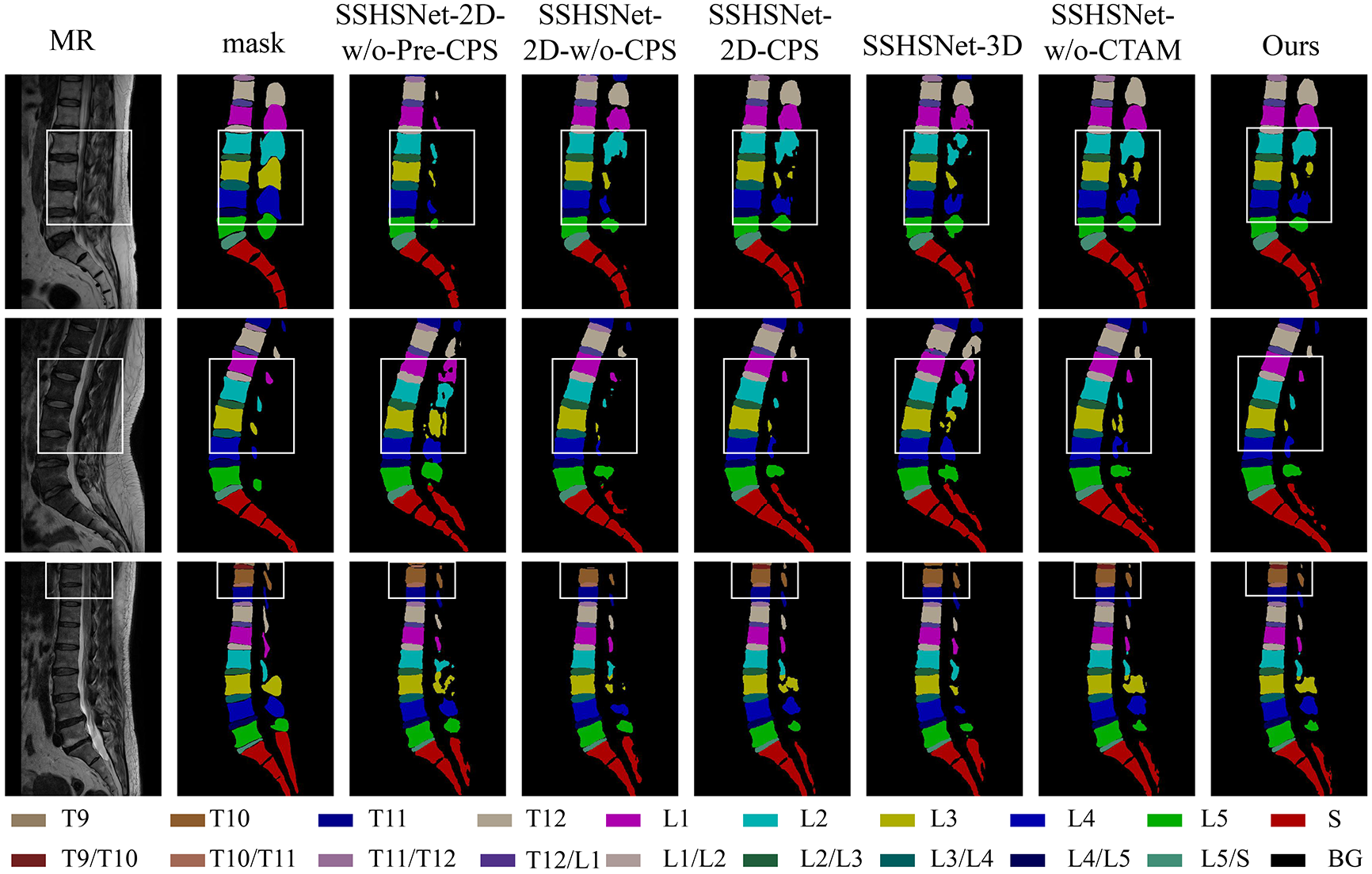}
  \caption{ Visualization results of the proposed method and its variations, where each row is a sagittal slice of a subject, and “BG” represents background. }
  \label{fig7}
\end{figure}

The CTAM was ignored in the proposed method, i.e., SSHSNet-w/o-CTAM, to assess the effectiveness of the proposed CTAM module on the performance of the proposed method. Compared with the SSHSNet-w/o-CTAM method, the proposed method shows increased segmentation accuracy (Fig. \ref{fig6} and Table \ref{t1}). This finding demonstrates that informative representation can be achieved by using CTAM for feature fusion, and good segmentation can be obtained. Compared with SSHSNet-w/o-CTAM, the proposed SSHSNet exhibits slightly improved segmentation performance (Fig. \ref{fig6}). However, the DSC values of T9 and T9/T10 are increased considerably by using the proposed SSHSNet (Table \ref{t1}), indicating that the proposed method can segment some small VBs and IVDs with an imbalance of category. The last two columns in Fig. \ref{fig7} also show that no remarkable difference is observed between the segmentation results of SSHSNet-w/o-CTAM and SSHSNet. However, SSHSNet is better at details than SSHSNet-w/o-CTAM, such as L3 segmentation in the second row. This result may contribute to the informative representation achieved by using CTAM for feature fusion in SSHSNet.

\subsection{Comparison with Other Methods}
\begin{figure}
  \centering
  \includegraphics[width=8cm]{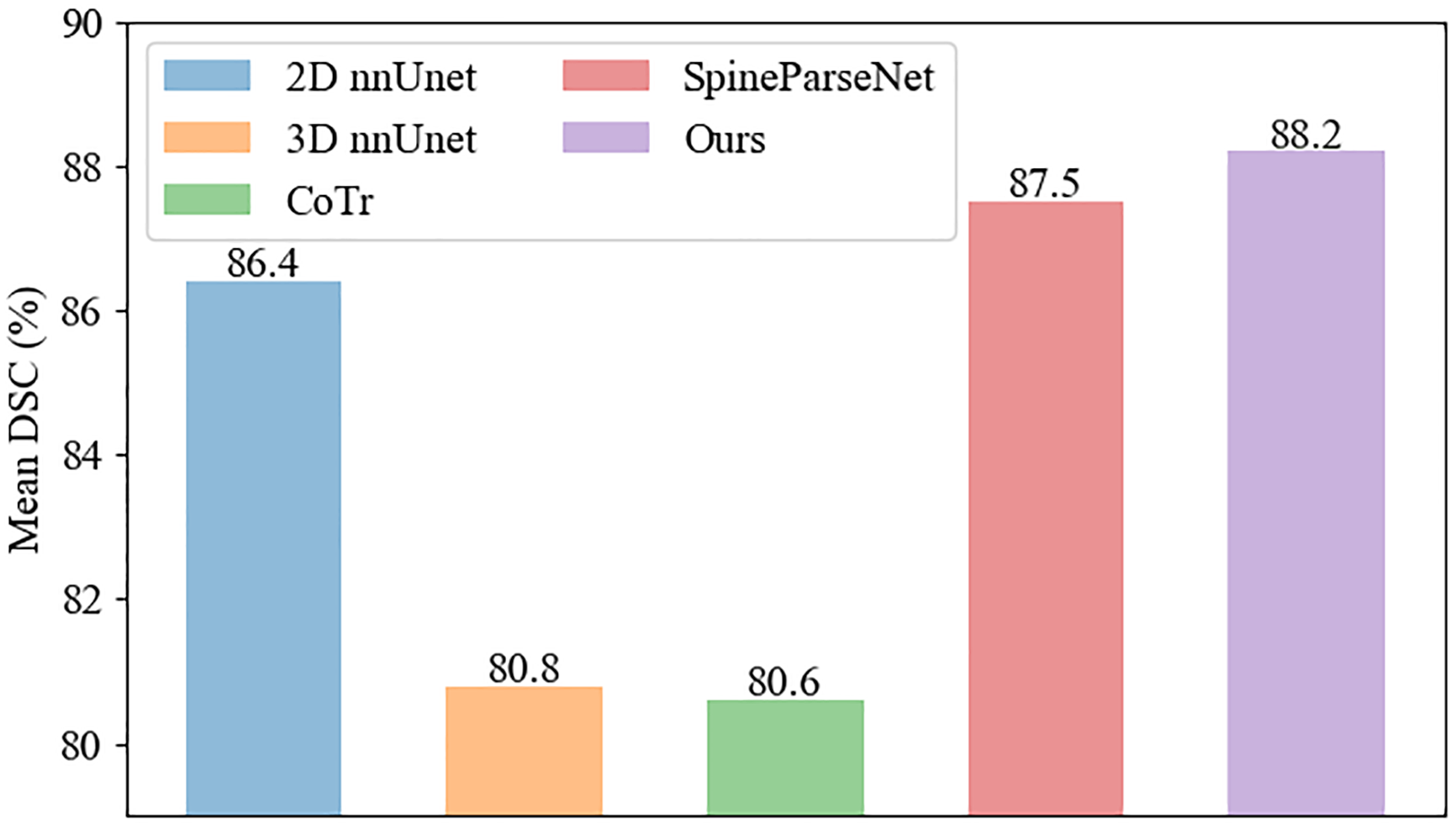}
  \caption{Overall mean DSC (\%) of different segmentation methods on testing set.}
  \label{fig8}
\end{figure}

We compared the proposed method with three other state-of-the-art methods, which include 2D and 3D nnUnet \cite{r24}, CoTr \cite{r25}, and SpineParseNet \cite{r2}, to further evaluate the performance of the proposed method on spine segmentation. The following results can be observed from Fig. \ref{fig8} and Table \ref{t2}. The segmentation accuracy of 3D nnUnet is similar to that of CoTr. The segmentation performance achieved by using 2D nnUnet is better than that achieved by using 3D nnUnet and CoTr. The segmentation accuracy when two-stage methods (SpineParseNet and the proposed method) are used is higher than that when single 2D (2D nnUnet) or 3D (3D nnUnet and CoTr) methods are used. The proposed method achieves the highest overall mean DSC value among all compared methods. Moreover, the achieved mean DSC in the segmentation of spine structures with unbalanced data labels, such as T9 VB and T9/T10 IVD, using the proposed method is much higher than that using other compared methods. This observation indicates the effectiveness of the proposed method on spine segmentation.

	\begin{landscape}
\renewcommand\arraystretch{1.2}
\setlength\tabcolsep{3pt}
\begin{table*}
  \scriptsize
  \centering
  \caption{Mean DSC (\%) of different segmentation methods for individual VB and IVD segmentation on the testing set.}
  \label{t2}
  \begin{tabular}{llllllllllllllllllll}
\hline
              & S              & L5             & L4             & L3             & L2             & L1             & T12            & T11            & T10            & T9             & L5/S           & L4/L5          & L3/L4          & L2/L3          & L1/L2          & T12/L1         & T11/T12        & T10/T11        & T9/T10         \\ \hline
2D nnUnet     & 89.06          & 88.65          & 88.91          & 89.01          & 89.08          & 89.92          & 89.45          & 80.72          & 62.29          & 38.91          & 86.59          & 86.70          & 88.40          & 88.32          & 89.22          & 88.76          & 87.23          & 61.74          & 55.81          \\
3D nnUnet     & 88.59          & 87.70          & 88.35          & 87.51          & 87.01          & 87.81          & 87.08          & 74.85          & 31.30          & 10.07          & 85.94          & 86.30          & 87.09          & 86.67          & 87.47          & 86.94          & 85.53          & 50.23          & 25.05          \\
CoTr          & 88.21          & 87.31          & 88.12          & 88.00          & 87.49          & 86.96          & 85.41          & 73.31          & 29.89          & 8.38           & 85.98          & 86.53          & 88.05          & 87.75          & 88.08          & 85.83          & 84.62          & 49.27          & 18.65          \\
SpineParseNet & \textbf{89.22} & \textbf{89.63} & \textbf{89.64} & \textbf{89.87} & \textbf{90.08} & \textbf{90.70} & 89.53          & 82.63          & 60.15          & 36.62          & \textbf{87.74} & 88.03          & \textbf{89.98} & \textbf{90.01} & 90.43          & 89.77          & 88.10          & 69.83          & 48.73          \\
Ours          & 87.91          & 88.66          & 89.60          & 89.56          & 89.64          & 90.20          & \textbf{89.53} & \textbf{83.23} & \textbf{71.72} & \textbf{75.77} & 87.71          & \textbf{88.31} & 89.95          & 89.88          & \textbf{90.76} & \textbf{90.41} & \textbf{88.42} & \textbf{70.02} & \textbf{84.36} \\ \hline
\end{tabular}
\end{table*}
	\end{landscape}
\begin{figure}
  \centering
  \includegraphics[width=8.8cm]{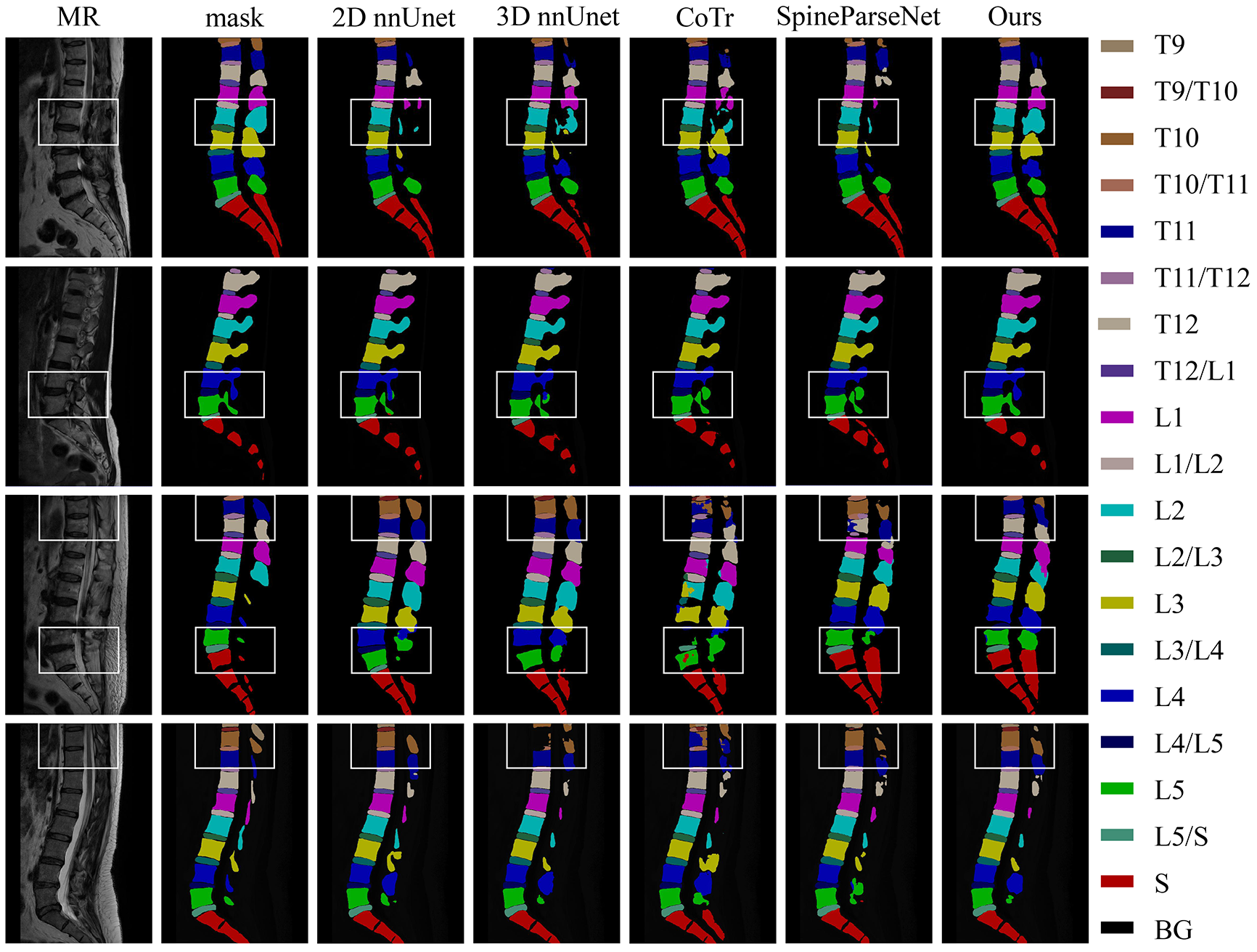}
  \caption{Visualization results of different methods, where each row is a sagittal slice of a subject, and “BG” represents background.}
  \label{fig9}
\end{figure}
The visualized segmentation results of our method and other state-of-the-art methods are shown in Fig. \ref{fig9}. The first row in Fig. \ref{fig9} shows that the vertebral arches are properly segmented by using the proposed method, indicating that the proposed method can alleviate the inferior segmentation caused by intensity inhomogeneity. The second row shows that the boundary between vertebral arches of L5 and L4 VB is blurred, resulting in the inferior segmentation in most advanced algorithms due to the partial volume effect. Compared with these methods, the proposed method can achieve appropriate segmentation for fuzzy bony structures. As shown in the upper white rectangle in the third row, the segmentation of the proposed method is more similar to the mask than that of the other methods, indicating that the proposed method can handle the issue of the inter-class similarity to some extent. Moreover, our algorithm is stable for L5 VB segmentation (the lower white rectangle in the third row). In comparison, the compared methods have few superfluous segmentations of other non-spinal structures, indicating the stability of our algorithms. The fourth row shows that most spine structures can be correctly identified by using the proposed method, and the segmentation performance achieved by using the proposed method is better than that achieved by using the other compared methods.

\section{discussion}
In this study, a novel two-stage method, named SSHSNet, is introduced to achieve accurate spine segmentation in volumetric MR images. In the first stage, 2D DeepLabv3+ networks combined with CPS are applied to explore global intra-slice information. Moreover, the coarse segmentation and the global intra-slice information provided in this stage can be used as prior knowledge for the next stage. In the second stage, a 3D DeepLabv3+ network is first adopted to exploit the inter-slice information. Then, a CTAM is used to combine intra- and inter-slice information effectively for fine spine segmentation. The proposed SSHSNet is evaluated on 215 spine MR images which contain normal and abnormal spine structures, provided by a publicly available dataset. The results show that the proposed SSHSNet has high accuracy in VB and IVD segmentation. Thus, the proposed method may provide a new way for automatic and accurate segmentation of multiple spine structures and has great potential in clinical spinal disease diagnoses and treatments.

Rich intra- and inter-slice information is included in a 3D spine MR images. Thus, the mean DSC value in SSHSNet-3D is higher than those in SSHSNet-2D-w/o-Pre-CPS under the same conditions of in-plane input resolution and network parameter initialized method (Fig. \ref{fig6} and Table \ref{t1}). However, compared with using SSHSNet-3D, using SSHSNet-2D-w/o-CPS in most VB and IVD segmentations achieves the coveted segmentation performance when the 2D network is trained on a pretrained model. This finding suggests that the training based on a pretrained model can help find an optimal solution in SSHSNet-2D-w/o-CPS. Moreover, compared with SSHSNet-2D-w/o-CPS and SSHSNet-3D, SSHSNet-2D-CPS considerably improves segmentation performance by incorporating CPS into the 2D network. In particular, the DSC values of some spine structures with unbalanced data labels, such as T9 VB and T9/T10 IVD, are higher in SSHSNet-2D-CPS than those in SSHSNet-2D-w/o-CPS and SSHSNet-3D. This scenario suggests that the data imbalance problem is alleviated, and model generalization ability is improved by incorporating CPS into the proposed method. A two-stage framework is presented in this study to embrace the advantages of inter-slice information contained in 3D spine MR images, improve the model generalization ability by using CPS, and reduce high computational memory cost in the whole 3D spine MR images by performing coarse-to-fine segmentation. As shown in Fig. \ref{fig6} and Table \ref{t1}, compared with SSHSNet-2D-w/o-CPS, SSHSNet-2D-CPS, and SSHSNet-3D, SSHSNet-w/o-CTAM improves segmentation performance, indicating the effectiveness of using a two-stage framework. However, compared with SSHSNet-2D-CPS, SSHSNet-w/o-CTAM exhibits some slightly decreased DSC values (Table \ref{t1}), such as the DSC values of T9 and T9/T10. This scenario may result from the redundant information generated by using simple 2D–3D feature concatenation in SSHSNet-w/o-CTAM. A CTAM is included in the proposed method to alleviate this issue by effectively combining 2D–3D features. Moreover, the DSC values of T9 and T9/T10 in the proposed SSHSNet are higher than those in SSHSNet-w/o-CTAM. Compared with SSHSNet-w/o-CTAM, the proposed SSHSNet improves the segmentation performance (Fig. \ref{fig7}), indicating the effectiveness of including CTAM into the proposed method.

Several state-of-the-art segmentation methods are performed in this study to assess the performance of the proposed method further. As shown in Fig. \ref{fig8} and Table \ref{t2}, the segmentation performance of 2D nnUnet is better than that of 3D nnUnet and CoTr, particularly for the segmentation of T9, T9/T10, and T10. The reason may be that only a part of the 3D MR images are used as inputs in 3D nnUnet and CoTr, and the global context information is lost in these two 3D methods. Additionally, compared with other segmentation methods, the proposed method achieves the highest overall mean DSC value and much higher mean DSC values in the segmentation of spine structures with unbalanced data labels, indicating the effectiveness of incorporating CPS into the proposed method for improvement of model generalization ability. Moreover, the visualized results (Fig. \ref{fig9}) show that the proposed method can achieve stable segmentation results in different cases, such as boundary blurring and medium inhomogeneity caused by MR anisotropy. This finding indicates the effectiveness of incorporating CTAM into the proposed method to provide a comprehensive spine description.

In this study, model generalization ability is improved by using CPS in the proposed method. However, the sample number with the spinal disease is only a small proportion of the entire dataset. Therefore, additional samples with different categories of spinal diseases will be collected to improve and evaluate the performance of the proposed method in future research. Moreover, a robust spine segmentation framework is proposed in this study, where SE-ResNext50 is used as the backbone. Some other advanced network structures, such as Swin Transformer \cite{r26}, can be introduced to the proposed method in future work to potentially extract rich image features and achieve excellent spine segmentation results.
\bibliographystyle{IEEEtran}
\bibliography{ref}
\end{document}